\documentclass[runningheads]{llncs}
\usepackage{graphicx}
\usepackage{booktabs}
\usepackage{multirow}
\usepackage{subfig}
\usepackage{hyperref}
\usepackage{pifont}
\usepackage{amsmath}
\usepackage{xcolor}
\usepackage{amssymb}
\usepackage[scr=dutchcal]{mathalpha}
\usepackage[labelfont=bf,labelsep=period]{caption}

\begin{document}

\title{Demographic Aware Probabilistic Medical Knowledge Graph Embeddings of Electronic Medical Records}
\titlerunning{DARLING: Demographic Aware Probabilistic Medical KGE}
%
\author{
Aynur Guluzade
\and Endri Kacupaj
\and Maria Maleshkova
}
\authorrunning{Guluzade et al.}

\institute{
    University of Bonn, Bonn, Germany\\
    \email{s6aygulu@uni-bonn.de}\\
    \email{\{kacupaj,maleshkova\}@cs.uni-bonn.de}
}

\maketitle
%

%
%
\begin{abstract}
Medical knowledge graphs (KGs) constructed from Electronic Medical Records (EMR) contain abundant information about patients and medical entities. The utilization of KG embedding models on these data has proven to be efficient for different medical tasks. However, existing models do not properly incorporate patient demographics and most of them ignore the probabilistic features of the medical KG. In this paper, we propose DARLING (\textbf{D}emographic \textbf{A}ware p\textbf{R}obabi\textbf{L}istic med\textbf{I}cal k\textbf{N}owledge embeddin\textbf{G}), a demographic-aware medical KG embedding framework that explicitly incorporates demographics in the medical entities space by associating patient demographics with a corresponding hyperplane. Our framework leverages the probabilistic features within the medical entities for learning their representations through demographic guidance. We evaluate DARLING through link prediction for treatments and medicines, on a medical KG constructed from EMR data, and illustrate its superior performance compared to existing KG embedding models.

\end{abstract}

\section{Introduction}
In recent years, knowledge graphs (KGs) have been established for medical assistance as the underlying core component of clinical decision support systems (CDSSs)~\cite{8923457,sherimon2016ontodiabetic} and self-diagnostic symptom checkers~\cite{Rotmensch2017}.
Those KGs are often extracted from sources such as Electronic Medical Records (EMR) and store clinical information into a set of triples for representing medical entities as nodes and relations as the edges between them. Medical KG-based applications have been reported in different scenarios, such as treatment recommendations~\cite{8890434}, medicine recommendations~\cite{GONG2021100174}, and drug-to-drug similarity measurements~\cite{Celebi2019}.
Those applications usually are performed through a link prediction process and can be divided into two steps: 1) learn embeddings of medical entities and relations, 2) make predictions/recommendations according to these embeddings.

Several approaches rely on medical KG embeddings for recommendation tasks, work from Gong et al.~\cite{GONG2021100174} proposed a medicine recommendation framework that embeds medical entities such as diseases, medicines, patients, and their corresponding relations into a shared lower dimensional space. The authors use the embeddings to decompose the task into a link prediction process while considering the patient’s diagnosis and adverse drug reactions. Chen et al.~\cite{8890434} proposed a framework operating on medical KG for thyroid treatment recommendation with cold start based on TransD~\cite{ji2015transD} and network embeddings with a hierarchical structure.

While recent research~\cite{Wanyan2020Graph,Li2020PrTransX,GONG2021100174,8890434,Celebi2019} employs traditional KG embedding methods \cite{Bordes2013transE,Wang2014transH,Lin2015transR,ji2015transD} as a first step for representing patients and medical entities, it lacks the consideration of demographic meta-data. However, such information is very advantageous and even necessary for medical tasks. 
Several works~\cite{Huang2017demographics,Ihra2012demographics,Standish2001Demographics,Qin2003demographics} have investigated and shown the importance and development of demographics in different medical tasks and challenges.
Considering this, we argue that incorporating demographics as part of medical KGs allows us to retain patients' generic information and construct more accurate representations for medical entities.

Hence, in this paper, we propose DARLING (\textbf{D}emographic \textbf{A}ware p\textbf{R}obabi\textbf{L}\allowbreak-istic med\textbf{I}cal k\textbf{N}owledge embeddin\textbf{G}) -- the first demographic-aware medical KG embedding framework that explicitly incorporates demographics in the medical entities space by associating patient demographics with a corresponding hyperplane. Our framework leverages probabilistic features within medical entities for learning their representations through demographic guidance. We evaluate DARLING on link prediction for treatments and medicines, where it achieves improved results compared to multiple existing KG embedding models on standard metrics. Furthermore, for evaluation purposes, we construct a medical KG from the MIMIC-III~\cite{johnson2016mimic} data, which comprises medical elements relating to patient admissions, such as demographics, disease diagnosis, treatment procedures, etc. The medical KG contains diseases, treatments and medicines, where our construction method automatically links all extracted entities with existing biomedical knowledge graphs, including ICD-9~\cite{Schriml2012icd9} ontology and DrugBank~\cite{Wishart2017DrugBank}. With our work we make the following key contributions to the state of the art:

\begin{itemize}
    \item We propose DARLING, the first demographic-aware framework for learning probabilistic medical KG embeddings.
    \item We provide a method to construct a medical KG with demographics meta-data that links all extracted entities with existing biomedical knowledge graphs.
    \item We demonstrate DARLING’s effectiveness through extensive experiments and show its superior performance via link prediction on treatments and medicines. We also illustrate the sensitivity of different demographic categories in our framework. 
\end{itemize}

\noindent To facilitate reproducibility and reuse our framework implementation, alongside the medical knowledge graph construction method, the results are also publicly available\footnote{\url{https://github.com/AynurGuluzade/DARLING}}.

The rest of the paper is structured as follows:
Section~\ref{sec:related_work} summarises the related work and Section~\ref{sec:approach} presents the proposed DARLING framework. Section~\ref{sec:experiments_evaluation} describes the experiments, including the experimental setup and the evaluation. Finally, we conclude in Section~\ref{sec:conclusion}.

\section{Related Work}\label{sec:related_work}
Our work lies at the intersection of medical KG embeddings and graph-based embedding approaches that employ patient demographics. In this section, we describe previous efforts and refer to different approaches.

Current efforts on knowledge graphs have concentrated on automatic knowledge base completion and population. Multiple KGs~\cite{ernst2015knowlife,dumontier2014bio2rdf} have been constructed from vast volumes of medical databases over the last years. Medical KGs contain medical facts of medicines and diseases and provide a pathway for medical discovery and applications, such as effective treatment and medicine recommendation. Unfortunately, such medical KGs suffer from severe data incompleteness problems, which impedes their application in clinical medicine. Celebi et al.~\cite{Celebi2019} proposed a KG embedding approach for drug-to-drug interaction prediction in a realistic scenario. Hettige et al.~\cite{hettige2020MedGraph} proposed an EMR embedding framework that introduces a graph-based data structure to capture visit-code associations in an attributed bipartite graph and the temporal sequencing visits through a point process.
Moreover, Choi et al.~\cite{Choi2016Med2Vec} proposed another approach that learns the representations for both medical codes and visits from large datasets.
Both works~\cite{hettige2020MedGraph,Choi2016Med2Vec} consider patient demographics on medical codes; however, they only focus on visit-code embeddings and do not directly harness medical KG embeddings or any medical task. Hence, specific medical information is insufficiently tailored. To the best of our knowledge, none of the existing approaches employs demographics as an essential  factor for medical KG embeddings. 

\section{DARLING}\label{sec:approach}

\begin{figure}[t!]
\centering
\includegraphics[width=0.8\textwidth]{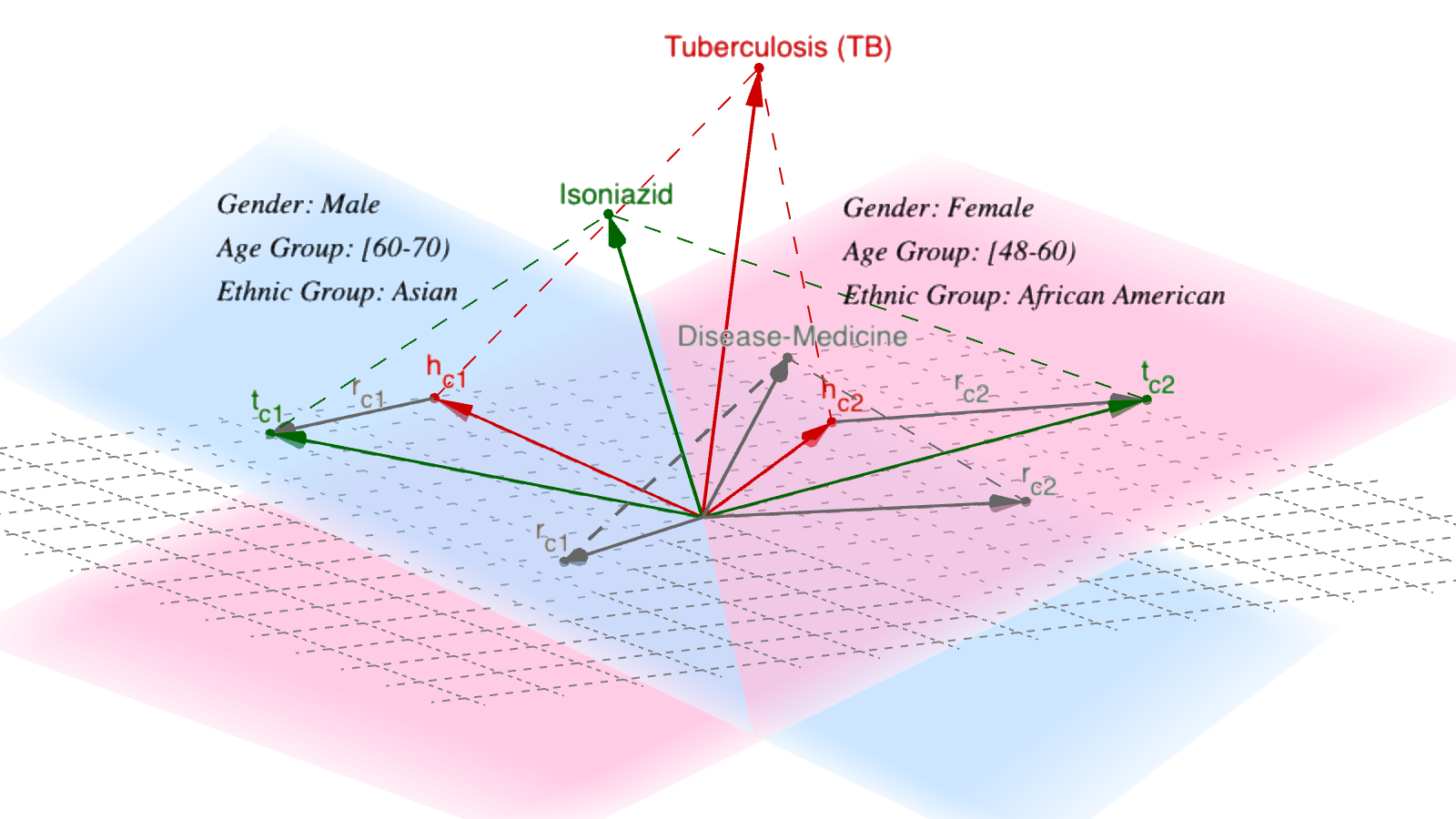}
\caption{The medical triple (Tuberculosis (TB), Disease-Medicine, Isoniazid) is valid for both demographic sets $c_1$ and $c_2$. $\boldsymbol{h}_{(c1)}, \boldsymbol{r}_{(c1)}$ and $\boldsymbol{t}_{(c1)}$ are the projections of the triple onto the demographic hyperplane $c_1$. As we can observe, the hyperplanes encode demographic information such as gender, age group, and ethnic group. DARLING learns demographic-aware representations of medical entities and relations by minimizing the translational distance $\|\boldsymbol{h}_c + \boldsymbol{r}_c - \boldsymbol{t}_c\|_{2}$ based on the triple probability score.}
\label{fig:darling_translation}
\end{figure}

This section presents a detailed description of DARLING (Figure~\ref{fig:darling_translation}).
In this work, we use a boldface lower-case letter $\boldsymbol{x}$ to denote a vector, $\|\boldsymbol{x}\|_p$ to represent its $l_p$ norm and $d$ for the embedding dimension.
Please refer to the appendix for the background of our framework, detailed experiment results, etc.

\subsection{Probabilistic Medical Knowledge Graph with Demographics}
A KG can be denoted as a set of triples $\mathcal{K} \subseteq \mathcal{E} \times \mathcal{R} \times \mathcal{E}$ where $\mathcal{E}$ and $\mathcal{R}$ are the set of entities and relations respectively. More precisely, the KG comprises triples $(h, r, t) \in \mathcal{K}$ in which $h, t \in \mathcal{E}$ represent a triples’ respective head and tail entities and $r \in \mathcal{R}$ represents its relation. The direction of a relationship indicates the roles of the entities, i.e., head or tail entity. In our scenario, we consider medical entities, extracted from EMR data, such as disease diagnosis, treatment procedures, and medicines; therefore $\mathcal{E} \subseteq \mathcal{D} \cup \mathcal{P} \cup \mathcal{M}$, where $\mathcal{D}$ is the set of diagnosis, $\mathcal{P}$ is the set of procedures and $\mathcal{M}$ is the set of medicines. In particular, we construct triples with $h \in \mathcal{D}$ and $t \in \mathcal{P} \cup \mathcal{M}$. Hence, our medical KG is denoted as $\mathcal{K} \subseteq \mathcal{D} \times \mathcal{R} \times \mathcal{P} \cup \mathcal{M}$.

We incorporate patient demographic meta-data such as gender, age group and ethnic group, by adding a new demographic dimension to the KG triples. We define the demographic set as $\mathcal{C} \subseteq \mathcal{G} \times \mathcal{A} \times \mathcal{T}$, where $\mathcal{G}$ is the set of genders, $\mathcal{A}$ the set of age groups and $\mathcal{T}$ the set of ethnic groups. Our medical KG contains quadruples $(h, r, t, c) \in \mathcal{K}$ where $c \in \mathcal{C}$ is a demographic set for which the corresponding medical triple $(h, r, t)$ holds. We aim to incorporate the demographic meta-fact $c$ directly into our learning algorithm, to learn demographic-aware embeddings of the KG elements.

Moreover, inspired by~\cite{Li2020PrTransX}, we strengthen each quadruple $(h, r, t, c)$ existence by introducing a statistical probability that indicates how likely is the particular triple $(h, r, t)$ to appear with the demographic set $c$. Specifically, we associate each quadruple with the probability $\mathscr{p}(h, r, t, c)$ which is calculated as, $\mathscr{p}(h, r, t, c) = \mathcal{N}_{(h, r, t, c)} / \mathcal{N}_{h}$, where $\mathcal{N}_{(h, r, t, c)}$ is the number of EMR admissions that the quadruple $(h, r, t, c)$ was extracted, while $\mathcal{N}_{h}$ is the number of admissions that contain the medical entity $h$. In contrast to~\cite{Li2020PrTransX}, our probability value is calculated by considering also the demographic set $c$ and not only the triple $(h, r, t).$

\subsection{Demographic-Guided Translation}
DARLING operates as an interaction model $f: \mathcal{D} \times \mathcal{R} \times \mathcal{P} \cup \mathcal{M} \to \mathbb{R}$ that computes a real-value score representing a medical KG quadruple's plausibility, given the embeddings for the entities, relations, and demographic sets. In our approach, we want the medical entity to have a distributed representation associated with different demographic sets. We achieve that by representing a demographic set as a hyperplane i.e., we will have $|\mathcal{C}|$ number of different hyperplanes represented by normal vectors $\boldsymbol{w}_1, \boldsymbol{w}_2,\dots , \boldsymbol{w}_{|\mathcal{C}|}$, where $|\mathcal{C}|$ denotes the total number of demographic sets. Therefore, we attempt to segregate the space into different demographic zones with the help of the hyperplanes. In this way, medical triples valid with demographic set $c$ are projected onto the demographic-specific hyperplane $\boldsymbol{w}_c \in \mathbb{R}^{d}$, where their translational distance (in our case similar to~\cite{Bordes2013transE}) is minimized.
Figure~\ref{fig:darling_translation} illustrates an example where the medical triple (h, r, t) is valid for both demographic sets $c_1$ and $c_2$. Hence it is projected onto the hyperplanes corresponding to those demographic sets. Using the triple embeddings $\boldsymbol{h}, \boldsymbol{r}, \boldsymbol{t} \in \mathbb{R}^d$, the projected representations on $\boldsymbol{w}_c$ are computed as:

\begin{equation}
\begin{split}
    \boldsymbol{h}_c &= \boldsymbol{h} - \boldsymbol{w}^{\top}_c \boldsymbol{h} \boldsymbol{w}_c, \\
    \boldsymbol{r}_c &= \boldsymbol{r} - \boldsymbol{w}^{\top}_c \boldsymbol{r} \boldsymbol{w}_c, \\
    \boldsymbol{t}_c &= \boldsymbol{t} - \boldsymbol{w}^{\top}_c \boldsymbol{t} \boldsymbol{w}_c, \\
\end{split}
\end{equation}

\noindent where $\boldsymbol{h}_c, \boldsymbol{r}_c, \boldsymbol{t}_c \in \mathbb{R}^d$. Consequently, we expect that a positive triple, valid with demographic set $c$, will have the mapping as $\boldsymbol{h}_c + \boldsymbol{r}_c \approx \boldsymbol{t}_c$. Accordingly, our scoring function is defined as:

\begin{equation}
    f_c(h, r, t) = \|\boldsymbol{h}_c + \boldsymbol{r}_c - \boldsymbol{t}_c\|_{p},
\end{equation}
\noindent where $p \in \{1, 2\}$ is a hyper-parameter.

Alongside the entity and relation embeddings, we also learn ${\{\boldsymbol{w}_c}\}^{|C|}_{c=1}$ for each demographic set $c$. Furthermore, by projecting the triple onto its demographic hyperplane, we incorporate demographic knowledge into the entity and relation embeddings, i.e., the same distributed representation will have a different role in different demographic sets.

\subsection{Optimization through Probability Score}
DARLING employs a margin-based pairwise ranking loss to differentiate between correct/positive and incorrect/negative triples. The negative triples are obtained by corrupting the positive one; thus, the pairs often share common head or tail entities and relations. Formally, we aim to minimize the following loss function:

\begin{equation}
    \mathscr{L} = \sum_{c\in [C]} \sum_{x\in \mathscr{D}^{+}_{c}} \sum_{y\in \mathscr{D}^{-}_{c}} \max(0,g_c(x) - g_c(y) + \gamma),
\end{equation}

\noindent with respect to the entity, relation and demographic set vectors. $\mathscr{D}^{+}_{c}$ is the set of valid triples with demographic set $c$, the negative triples are drawn from the
set $\mathscr{D}^{-}_{c}$, and $\gamma$ is a margin separating correct and incorrect triples.

Unlike existing approaches that employ pairwise ranking loss, DARLING does not directly use the score function $f_{c}(h, r, t)$. Instead, we utilize probability values $\mathscr{p}$ for estimating the scores for optimization.
First, we introduce a mapping function or probability score function $f_{\mathscr{p}}$ that allows us to map each quadruple probability value $\mathscr{p}(h, r, t, c)$ into a score. This function is defined as:

\begin{equation}
    f_{\mathscr{p}}(h, r, t, c) = \lambda \ln{\mathscr{p}(h, r, t, c)^{-1}},
\end{equation}

\noindent where $\lambda$ is a scaling factor. To avoid a $0$ denominator for negative quadruples, we introduce a constant probability value $\varepsilon_n > 0$. Furthermore, we set a minimum probability value for positive quadruples as $\varepsilon_p$, where $\varepsilon_p > \varepsilon_n$. Second, we define the function $g_c()$ as the absolute difference of the probability score $f_{\mathscr{p}}$ and the score function $f_c$. Formally, this is described as:

\begin{equation}
    g_c(h, r, t) = |f_{\mathscr{p}}(h, r, t, c) - f_c(h, r, t)|.
\end{equation}

In this way, the DARLING optimization process allows us to learn representations that would satisfy each quadruple probability value. Specifically, the quadruple entities with a high probability value will have representations closer in space compared to those with a low probability.

\section{Experiments}\label{sec:experiments_evaluation}
\subsection{Datasets and Medical Knowledge Graph Construction}
\begin{figure}[t!]
\centering
\includegraphics[width=0.8\textwidth]{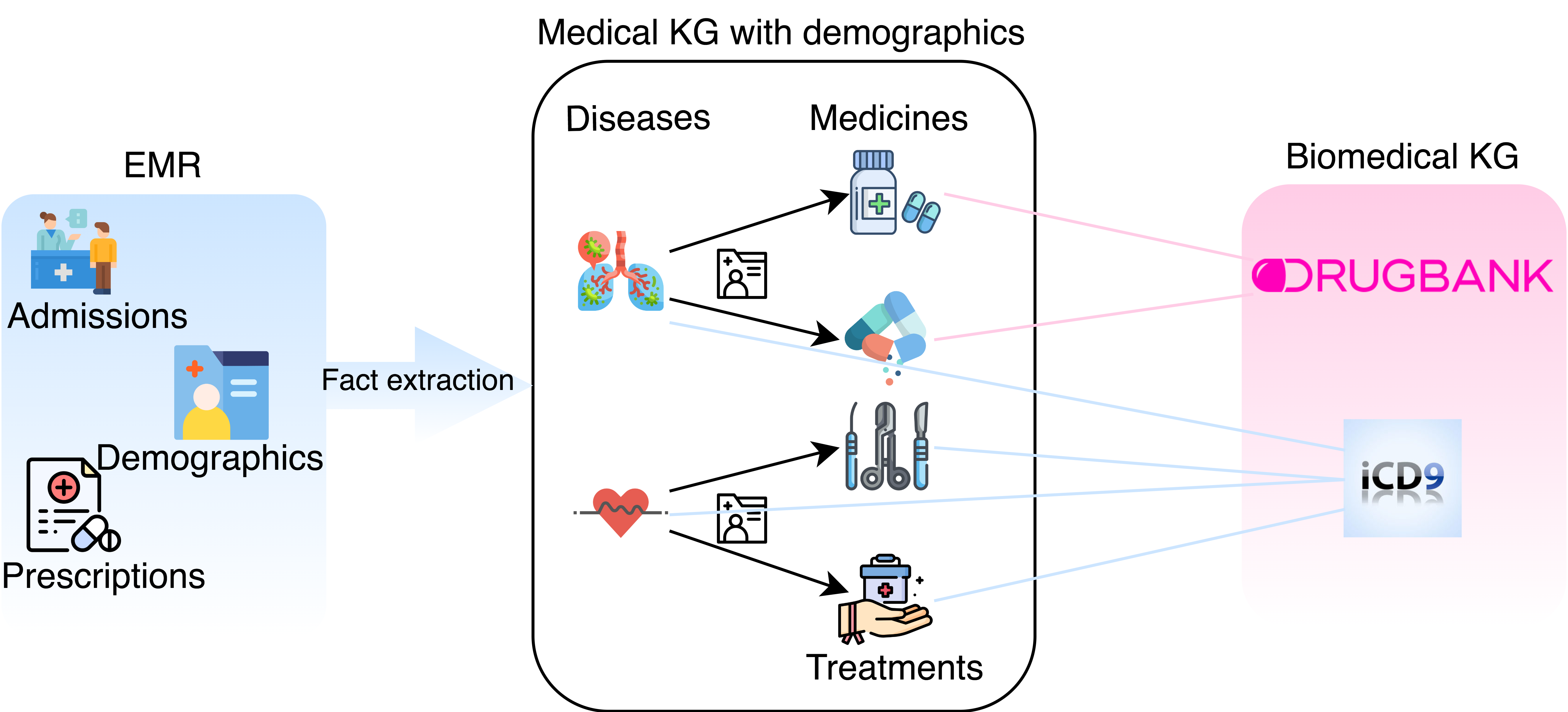}
\caption{Overview of our Medical KG construction process.}
\label{fig:kg_construction}
\end{figure}

We perform experiments on a real EMR dataset -- MIMIC-III~\cite{johnson2016mimic}, and two biomedical knowledge graphs, DrugBank~\cite{Wishart2017DrugBank} and the ICD-9~\cite{Schriml2012icd9} ontology. MIMIC-III (Medical Information Mart for Intensive Care) comprises information related to patients admitted to critical care units at a large tertiary care hospital. The dataset contains distinct information about $46,520$ patients, $58,976$ admissions, and $1,517,702$ prescription records associated with $6,985$ diagnosis, $2,032$ procedures and $4,525$ medicines. For our work, we extract patient demographics, disease diagnosis, treatment procedures, and medicines. We link the extracted medicines to DrugBank, which is a bioinformatics resource that consists of medicine-related entities. The DrugBank KG contains $8,054$ medicines, $4,038$ other related entities (e.g., protein or drug targets) and $21$ relationships.
Moreover, we link extracted diseases and treatments with ICD-9 ontology (International Classification of Diseases, Ninth Revision) which contains $13,000$ international standard codes of diagnosis and procedures. We connect MIMIC-III, DrugBank, and the ICD-9 ontology by constructing the medical KG~(c.f. Figure~\ref{fig:kg_construction}). 
Our medical KG includes $9,289$ distinct entities and two types of relations -- Disease\_to\_Treatment and Disease\_to\_Medicine. Regarding demographics, we end up with $79$ different demographic set combinations (gender, age group and ethnic group). Finally, our KG contains $126,141$ distinct quadruples, $100,912$ of which we use for training, $10,091$ for validation and $15,138$ for testing. Table~\ref{tab:kg_details} gives details on the constructed KG.

\begin{table}[t!]
\centering
\def\arraystretch{1.2}
\caption{Medical KG number of entities, relations and demographics.}
\label{tab:kg_details}
\begin{tabular}{lc|lc|lc}
\toprule
\multicolumn{2}{c|}{\textbf{Entities}} & \multicolumn{2}{c|}{\textbf{Relations}} & \multicolumn{2}{c}{\textbf{Demographics}} \\ \midrule
\#Disease & 6,968  & \#Disease\_to\_Treatment & 58,225 & \#Gender & 2 \\
\#Treatment & 1,475 & \#Disease\_to\_Medicine  & 67,916 & \#Age group & 6 \\
\#Medicine & 846  & & & \#Ethnic group & 7 \\ \midrule
\#Total & 9,289 & \#Total & 126,141 & \#Total (sets) & 79 \\ 
\bottomrule
\end{tabular}
\end{table}

\subsection{Experimental Setup}
\paragraph{\textbf{Model Configurations}}
For the experiments, we selected Adam optimizer, and we employ batch sizes of $b = \{ 128, 256, 512 \}$, embedding dimensions of $d=\{128, 256, 512\}$, learning rates $lr=\{0.01, 0.001, 0.0001\}$, a margin $\gamma=1$ and $p=2$ for the scoring function. Furthermore, for the probabilistic hyper-parameters $\lambda$, $e_p$ and $e_n$ we use the values of $10^{-2}$, $10^{-4}$, and $10^{-15}$ respectively. We train DARLING for $100$ epochs and select the best state by the corresponding lowest mean rank on the validation set. 

\paragraph{\textbf{Models for Comparison}}
For evaluating the performance of our framework, we compare against the following methods:

\begin{itemize}
    \item \textbf{TransE}~\cite{Bordes2013transE}~is a simple but effective translation-based model. A major advantage of TransE is its computational efficiency, which enables its use for large-scale KGs.
    \item \textbf{TransH}~\cite{Wang2014transH}~is an extension of TransE where each relation is represented by a hyperplane. Our proposed framework, DARLING also adjusts TransE in a similar way by treating the demographic sets as hyperplanes.
    \item \textbf{TransR}~\cite{Lin2015transR}~explicitly considers entities and relations as different objects and therefore represents them in different vector spaces.
    \item \textbf{TransD}~\cite{ji2015transD} is similar to TransR, however, instead of performing the same relation-specific projection for all entity embeddings, entity-relation-specific projection matrices are constructed.
    \item \textbf{PrTransE \& PrTransH}~\cite{Li2020PrTransX}~are extensions of TransE and TransH, which introduce triple probabilities. Our optimization via probability score was inspired by this work, however, our approach differs considerably in multiple aspects. 
\end{itemize}

\subsection{Results}\label{sec:results}
We apply the following frequently used metrics to summarize the overall performance: 1) Mean rank (MR): which represents the average rank of the test triples, where smaller values indicate better performance. 2) Hits@K: which denotes the ratio of the test triples that have been ranked among the top-$k$ triples, where larger values indicate better performance. We report results for $k = \{3, 10\}$.

\begin{table}[t!]
\centering
\def\arraystretch{1.1}
\caption{Results on link prediction for treatments and medicines. DARLING outperforms all other methods.}
\label{tab:results}
\begin{tabular}{l|c|cc|c|cc}
\toprule
\textbf{Task} & \multicolumn{3}{c|}{\textbf{Disease-Treatment}} & \multicolumn{3}{c}{\textbf{Disease-Medicine}} \\ \midrule
Methods & Mean Rank & Hits@3 & Hits@10 & Mean Rank & Hits@3 & Hits@10 \\ \midrule
TransE~\cite{Bordes2013transE} & 73.94 & 15.71\% & 47.40\% & 27.04 & 15.89\% & 54.33\% \\
TransH~\cite{Wang2014transH} & 75.56 & 16.51\% & 48.60\% & 27.71 & 16.23\% & 55.46\% \\
TransR~\cite{Lin2015transR} & 115.12 & 12.64\% & 30.34\% & 45.74 & 14.33\% & 39.16\% \\
TransD~\cite{ji2015transD} & 84.66 & 17.34\% & 47.64\% & 33.51 & 15.97\% & 55.76\% \\
PrTransE~\cite{Li2020PrTransX} & 69.69 & 16.29\% & 47.21\% & 27.51 & 15.45\% & 54.80\% \\
PrTransH~\cite{Li2020PrTransX} & 69.01 & 16.89\% & 47.25\% & 26.71 & 16.14\% & 55.73\% \\ \midrule
\textbf{DARLING (ours)} & \textbf{64.65} & \textbf{22.71\%} & \textbf{52.19\%} & \textbf{22.86} & \textbf{26.90\%} & \textbf{61.73\%} \\
\bottomrule
\end{tabular}
\end{table}

\begin{figure}[h!]
\centering
\includegraphics[width=0.9\textwidth]{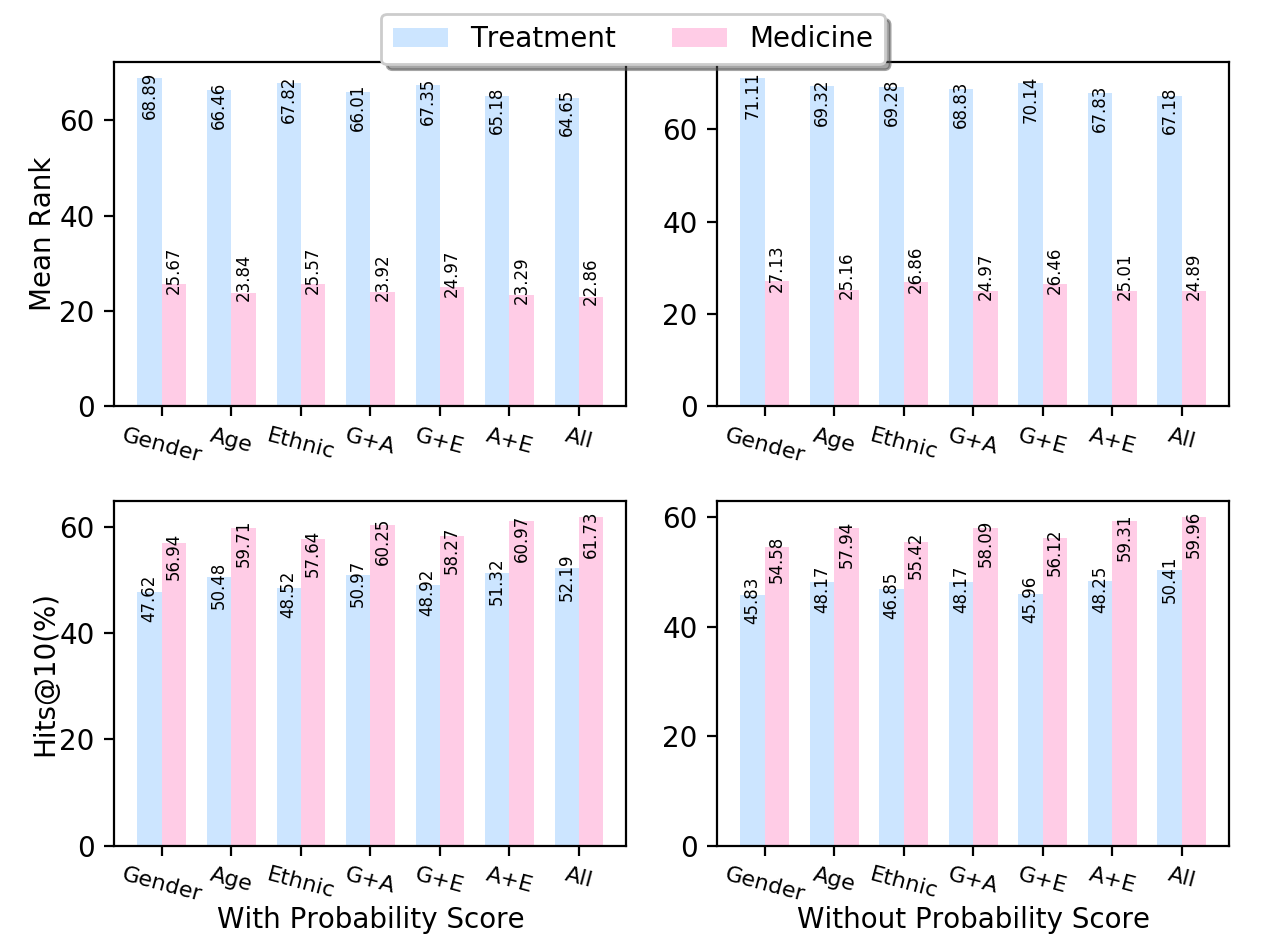}
\caption{DARLING results based on different demographic combinations with and without probability score.}
\label{fig:demo_sensitivity}
\end{figure}

Table~\ref{tab:results} illustrates the results for treatment and medicine prediction from our constructed medical KG.  As indicated, DARLING outperforms all other traditional KG embedding approaches on both mean rank and Hits@K. More precisely, for Hits@3 and Hits@10, our framework has an absolute difference of at least $5$ points from the baselines, while for Hits@3 on medicine prediction, this difference is increased to $10$ absolute points. DARLING also has lower mean rank compared to all other baselines.  The improved performance arises from DARLING's ability to generate more accurate embeddings of the medical entities. The demographic-based hyperplanes that our framework produces allow it to categorize diseases, treatments and medicines into ``subspaces'' where it learns their embeddings. Furthermore, the quadruple probability score provides some weighting for the distance between the entities. In contrast, all the baselines project the medical entities in one or two spaces (since the number of relations is two). As demonstrated by our results, this is not sufficient to represent all entities accurately.

\subsection{Demographic and Probability Score Sensitivity}\label{sec:sensitivity}
We perform experiments to identify the demographic and probability score sensitivity of our framework. In particular, we aim to recognize which demographic category (gender, age group, ethnic group) is more effective and whether the probability score impacts our results. To do so, we adjust DARLING to construct hyperplanes with all demographics individually and with all possible combinations. At the same time, we run the framework by including and excluding the probability scores.

Figure~\ref{fig:demo_sensitivity} illustrates the results of this experiment for mean rank and Hits@10. Regarding the demographics, we obtain that ``Age group'' is the most prominent one since the results are better compared to ``Gender'' and ``Ethnic group''. We find this accurate since most diseases are highly correlated with patient age. Furthermore, we observe that with ``Ethnic group'' we obtain better results than with ``Gender''. When joining the categories ``Age group + Ethnic group (A+E)'' performs better than ``Gender + Age group (G+A)'' and ``Gender + Ethnic group (G+E)''. It is worth mentioning that even when employing one demographic category together with the probability score, we still acquire better results than any of the baseline methods. We obtain the highest possible results when we utilise all three demographic categories, as we also do for DARLING. Moreover, the results indicate a higher performance by 1-3 points when using the probability score. 

\section{Conclusions}\label{sec:conclusion}
In this article, we focus on medical knowledge graph embeddings of electronic medical records. We provide a demographic-aware embedding framework that explicitly incorporates demographics in the medical entities' space by associating patient demographics (gender, age, ethnicity) with a corresponding hyperplane. Our framework leverages probabilistic features of entities for learning their embeddings through demographic guidance. Furthermore, for evaluating our approach, we present a method to construct a medical KG from EMR data and automatically link all extracted entities with existing biomedical knowledge graphs (ICD-9 and DrugBank). We empirically show that our model achieves the best results in link prediction for treatments and medicines, compared to other traditional KG embedding approaches. Moreover, we perform a demographic sensitivity experiment and discover that age is the demographic category that significantly affects our high results. We also show the importance of probability score in our framework.
For future work, we intend to employ our framework as an embedding method for existing medical recommendation systems.

\vspace{-0.5em}

\bibliographystyle{splncs04}
\bibliography{main.bbl}

\appendix

\section*{Appendix}\label{sec:appendix}

\section{Background}
In this section, we review existing methods for knowledge graph representation learning~\cite{Bordes2013transE},~\cite{Wang2014transH}.
A KG is considered as a set of entities $\mathcal{E}$ and relations $\mathcal{R}$. The set of directed edges, $\mathscr{D}^{+}$ comprises triples $(h, r, t)$ where a direction of relation $r$  is from head $h$ to tail $t$ entity

TransE~\cite{Bordes2013transE} is a simple and efficient translational based distance model. It models the relation as a translation vector between head and tail entity vectors. For the given two entity vectors $\boldsymbol{h}$, $\boldsymbol{t} \in \mathbb{R}^{n}$, it maps the relation as translation vector $\boldsymbol{r} \in \mathbb{R}^{n}$,i.e., $\boldsymbol{h} + \boldsymbol{r} \approx \boldsymbol{t}$ for observed triple $\boldsymbol{h, r, t}$. Thus, the distance based scoring function is defined as:

\begin{equation}
    f(h, r, t) = \|\boldsymbol{h} + \boldsymbol{r} - \boldsymbol{t}\|_{l_1/l_2},
\end{equation}
\noindent where, $\|\cdot\|_{l_1/l_2}$ is the $l_1$ or $l_2$-norm of the difference vector. $f(h,r,t)$ will be minimized for plausible triples. A margin based pairwise ranking loss is used to differentiate between correct and incorrect triples by minimizing their TransE score difference. Formally, the loss function is defined:

\begin{equation}
    \sum_{x\in \mathscr{D}^{+}} \sum_{y\in \mathscr{D}^{-}}
    \max(0, f(x) -f(y) + \gamma),
\end{equation}

\noindent with respect to the entity and relation vectors. $\gamma$ is a margin hyperparameter. $\mathscr{D}^{+}$  stores only positive triples, i.e., observed triples in KG. $\mathscr{D}^{-}$ is the set with negative examples that are drawn randomly.

Despite its simplicity and efficiency, TransE cannot model 1-to-N, N-to-1, and N-to-N type of relations as it does not learn a distributed representation of entities. To tackle these flaws, TransH~\cite{Wang2014transH} was introduced to model a relation $r$ as a vector on a relation specific hyperplane, and project entities associated with it on the corresponding hyperplane for learning the entities' distributed representation.

\section{Negative Sampling}
For negative sampling, DARLING utilizes a uniform demographic agnostic approach, where it considers the set of all the triples that do not belong to the medical KG, irrespective of the demographic dimension. Formally, for the demographic set $c$, the negative samples are drawn from the set,

\begin{equation}
\begin{split}
    \mathscr{D}^{-}_{c} = &\{ (h^{'}, r, t)|h^{'} \in \mathcal{E},  (h^{'}, r, t) \notin \mathscr{D}^{+} \} \\ \cup &\{ (h, r, t^{'})|t^{'} \in \mathcal{E},  (h, r, t^{'}) \notin \mathscr{D}^{+} \}.
\end{split}
\end{equation}

\section{Demographic Statistics}
Table~\ref{tab:demo_stats} illustrates the number of unique patients that belong to each demographic category. We constructed our medical KG using data from 46,520 patients and 58,976 admissions related to them. The grouping of age values (years) was done by us, considering that we wanted to distribute the patients equally in different groups. The genders and ethnic groups are adopted from MIMIC-III data \cite{johnson2016mimic}.

\begin{table}[t!]
\caption{Demographic statistics for each category. Our medical KG contains data from 46,520 unique patients.}
\label{tab:demo_stats}
\vspace{1em}
\begin{tabular}[t]{l|c}
    \toprule
    \textbf{Gender} & \textbf{\#Patients} \\
    \midrule
    male & 26,121 \\
    female & 20,399 \\
    \bottomrule
\end{tabular}
\hfill
\begin{tabular}[t]{l|c}
    \toprule
    \textbf{Age Group} & \textbf{\#Patients} \\
    \midrule
    {[}0-18) & 7,942 \\
    {[}18-48) & 7,005 \\
    {[}48-60) & 7,515 \\
    {[}60-70) & 7,860 \\
    {[}70-80) & 7,939 \\
    \textgreater{}= 80 & 8,259 \\
    \bottomrule   
\end{tabular}
\hfill
\begin{tabular}[t]{l|c}
    \toprule
    \textbf{Ethnic Group} & \textbf{\#Patients} \\
    \midrule
    white & 32,372 \\
    black & 3,871 \\
    asian & 1,690 \\
    hispanic & 1,642 \\
    native & 46 \\
    other & 1,489 \\
    unknown & 5,410 \\
    \bottomrule   
\end{tabular}
\end{table}

\section{Inference}
For inference, we describe how DARLING can be used for medical recommendation tasks through a link prediction process. Given a query patient with demographic set $\mathscr{c} \in \mathcal{C}$ (gender, age, ethnicity) and the query disease diagnosis $\mathscr{d} \in \mathcal{D}$, we use DARLING to project the disease $\mathscr{d}$ into the hyperplane $w_{\mathscr{c}}$ and recommend top-$k$ treatments and medicines. More precisely, given a query $\mathscr{q}=(\mathscr{c},\mathscr{d})$, for each treatment procedure $\forall \mathscr{p} \in \mathcal{P}$ and medicine $\forall \mathscr{m} \in \mathcal{M}$ we compute its triple score with $\mathscr{d}$ (i.e. $f_{\mathscr{c}}(\mathscr{d}, r, \mathscr{p}), f_{\mathscr{c}}(\mathscr{d}, r, \mathscr{m})$) on the demographic hyperplane $w_c$, and then select the treatment $\mathscr{p}$ and medicine $\mathscr{m}$ with the top-$k$ highest ranking scores as the recommendation.

\section{Detailed Results}
Table~\ref{tab:detailed_results} presents detailed results of our framework. In particular, we illustrate results using all possible demographic category combinations, and we further provide results by including and excluding the probability scores. At the same time, we present the results of all other baselines. As we can see, DARLING outperforms all baselines when using all demographic categories and including the probability scores

\begin{table}[t!]
\centering
\def\arraystretch{1.2}
\caption{Detailed results of our experiments.}
\label{tab:detailed_results}
\resizebox{\textwidth}{!}{%
\begin{tabular}{l|c|c|c|c|c|c|c|c}
\toprule
\textbf{Task} & \multicolumn{4}{c|}{Disease-Treatment} & \multicolumn{4}{c}{Disease-Medicine} \\ \midrule
Methods & \multicolumn{2}{c|}{Mean Rank} & \multicolumn{2}{c|}{Hits@10} & \multicolumn{2}{c|}{Mean Rank} & \multicolumn{2}{c}{Hits@10} \\ \midrule
TransE~\cite{Bordes2013transE} & \multicolumn{2}{c|}{73.94} & \multicolumn{2}{c|}{47.40\%} & \multicolumn{2}{c|}{27.04} & \multicolumn{2}{c}{54.33\%} \\
TransH~\cite{Wang2014transH} & \multicolumn{2}{c|}{75.56} & \multicolumn{2}{c|}{48.60\%} & \multicolumn{2}{c|}{27.71} & \multicolumn{2}{c}{55.46\%} \\
TransR~\cite{Lin2015transR} & \multicolumn{2}{c|}{115.12} & \multicolumn{2}{c|}{30.34\%} & \multicolumn{2}{c|}{45.74} & \multicolumn{2}{c}{39.16\%} \\
TransD~\cite{ji2015transD} & \multicolumn{2}{c|}{84.66} & \multicolumn{2}{c|}{47.64\%} & \multicolumn{2}{c|}{33.51} & \multicolumn{2}{c}{55.76\%} \\
PrTransE~\cite{Li2020PrTransX} & \multicolumn{2}{c|}{69.69} & \multicolumn{2}{c|}{47.21\%} & \multicolumn{2}{c|}{27.51} & \multicolumn{2}{c}{54.80\%} \\
PrTransH~\cite{Li2020PrTransX} & \multicolumn{2}{c|}{69.01} & \multicolumn{2}{c|}{47.25\%} & \multicolumn{2}{c|}{26.71} & \multicolumn{2}{c}{55.73\%} \\ \midrule
\textbf{Probability score} & with & without & with & without & with & without & with & without \\ \midrule
DARLING (Gender) & 68.89 & 71.11 & 47.62\% & 45.83\% & 25.67 & 27.13 & 56.94\% & 54.58\% \\
DARLING (Age) & 66.46 & 69.32 & 50.48\% & 48.17\% & 23.84 & 25.16 & 59.71\% & 57.94\% \\
DARLING (Ethnicity) & 67.82 & 69.28 & 48.52\% & 46.85\% & 25.57 & 26.86 & 57.64\% & 55.42\% \\
DARLING (G+A) & 66.01 & 68.83 & 50.97\% & 48.17\% & 23.92 & 24.97 & 60.25\% & 58.09\% \\
DARLING (G+E) & 67.35 & 70.14 & 48.92\% & 45.96\% & 24.97 & 26.46 & 58.27\% & 56.12\% \\
DARLING (A+E) & 65.18 & 67.83 & 51.32\% & 48.25\% & 23.29 & 25.01 & 60.97\% & 59.31\% \\ \midrule
\textbf{DARLING (all)} & \textbf{64.65} & \textbf{67.18} & \textbf{52.19\%} & \textbf{50.41\%} & \textbf{22.86} & \textbf{24.89} & \textbf{61.73\%} & \textbf{59.96\%} \\ \bottomrule
\end{tabular}%
}
\end{table}

\end{document}